\newif\ifmnras
\newif\ifemapj
\emapjfalse
\mnrasfalse
\ifmnras
\documentclass[useAMS,usenatbib,usedcolumn]{mnras}
\else
\ifemapj
\documentclass[apj,numberedappendix]{emulateapj}
\else
\documentclass[twocolumn,tighten]{aastex62}
\fi
\fi
\usepackage{graphics,epsf}
\usepackage{amsmath}                % American Mathematical Society package
\usepackage{amsfonts}               % American Mathematical Society fonts
\usepackage{amssymb}                % American Mathematical Society symbol
\usepackage{epsfig}                 % EPS figures
\usepackage{graphicx}
\usepackage{float}
\usepackage[para,online,flushleft]{threeparttable}

\newcommand{\km}{{~\rm km}}
\newcommand{\s}{{~\rm s}}

\newcommand{\yr}{{~\rm yr}}

\newcommand{\AU}{{~\rm AU}}

\ifmnras
\else
\ifemapj
	\newcommand{\nar}{{~\rm New Astronomy Reviews}}
	\newcommand{\na}{{~\rm New Astronomy}}
    \newcommand{\pasa}{{~\rm Publications of the Astronomical Society of Australia}}
\fi
\fi

\ifmnras
\title[Grazing envelope orbital evolution]{Orbital radius during the grazing envelope evolution}
\author[A. Abu-Backer, A. Gilkis, N. Soker]{Abedallah Abu-Backer$^{1}$\thanks{Contact e-mail: \href{abu-backer@campus.technion.ac.il}{abu-backer@campus.technion.ac.il}}, Avishai Gilkis$^{2}$\thanks{Contact e-mail: \href{agilkis@ast.cam.ac.uk}{agilkis@ast.cam.ac.uk}}, Noam Soker$^{1}$\thanks{Contact e-mail: \href{soker@physics.technion.ac.il}{soker@physics.technion.ac.il}}
\\
$^{1}$ Department of Physics, Technion -- Israel Institute of Technology, Haifa 3200003, Israel \\
$^{2}$ Institute of Astronomy, University of Cambridge, Madingley Road, Cambridge, CB3 0HA, UK 
}
\fi

\begin{document}

\ifmnras
\pagerange{\pageref{firstpage}--\pageref{lastpage}} \pubyear{2018}

\maketitle
\label{firstpage}
\else

\title{Orbital radius during the grazing envelope evolution}
\ifemapj
\author{Abedallah Abu-Backer\altaffilmark{1}, Avishai Gilkis\altaffilmark{2} \& Noam Soker\altaffilmark{1,3}}

\altaffiltext{1}{Department of Physics, Technion -- Israel Institute of Technology, Haifa 3200003, Israel; abu-backer@campus.technion.ac.il; soker@physics.technion.ac.il}
\altaffiltext{2}{Institute of Astronomy, University of Cambridge, Madingley Road, Cambridge, CB3 0HA, UK; agilkis@ast.cam.ac.uk}
\altaffiltext{3}{Guangdong Technion Israel Institute of Technology, Shantou 515069, Guangdong Province, China}
\else
\author{Abedallah Abu-Backer}
\affil{Department of Physics, Technion -- Israel Institute of Technology, Haifa 3200003, Israel}
\email{abu-backer@campus.technion.ac.il}

\author[0000-0001-8949-5131]{Avishai Gilkis}
\affil{Institute of Astronomy, University of Cambridge, Madingley Road, Cambridge, CB3 0HA, UK}
\email{agilkis@ast.cam.ac.uk}

\author[0000-0003-0375-8987]{Noam Soker}
\affil{Department of Physics, Technion -- Israel Institute of Technology, Haifa 3200003, Israel}
\affil{Guangdong Technion Israel Institute of Technology, Shantou 515069, Guangdong Province, China}
\email{soker@physics.technion.ac.il}

\shorttitle{Grazing envelope orbital evolution} 
\shortauthors{A. Abu-Backer et al.}
\fi
\fi

\begin{abstract}
We use the \textsc{binary} module of the \textsc{mesa} code to study the evolution of an evolved binary system where we assume that a main sequence companion removes the outskirts of the envelope of an asymptotic giant branch (AGB) star by launching jets, and explore the characteristics of this grazing envelope evolution (GEE). We base our assumption that jets launched by the secondary star remove a substantial fraction of the outskirts of the envelope of an AGB star on earlier hydrodynamical simulations. We find that in many cases that we study, but not in all cases, the binary system experiences the GEE rather than entering the common envelope phase, under our assumptions of jet-driven mass removal. To prevent the common envelope phase, we assume the secondary star may accrete a large amount of mass in a short time while avoiding rapid inflation, the feasibility of which requires further study. Because of our simplifying assumptions we cannot yet present the parameter space for the GEE. Although the incorporation of the GEE into population synthesis numerical codes requires further studies of the GEE, we conclude that analyses of population synthesis studies of evolved binary stars should include the GEE. 
\ifmnras
\else
\fi
\end{abstract}

\ifmnras
\fi

% ==========================================================
\section{INTRODUCTION}
\label{sec:intro}
% ==========================================================

Traditional studies of binary systems that are composed of a primary giant star and a more compact secondary star have considered two types of orbital evolution. In the first the compact object is outside the envelope and due to the mass loss process the orbital separation increases, and in the second type the compact object enters the envelope of the giant primary star and spirals-in, i.e., enters a common envelope phase. These two types of interaction have been implemented in population synthesis codes of stellar binary systems (e.g., \citealt{MoeDeMarco2006, Izzardetal2009, Leeetal2014, Toonenetal2014, Abateetal2015}), as well as in very detailed studies of the binary evolution that include the mass loss geometry, like the formation of a circumbinary disc (e.g., \citealt{Chenetal2017, Chenetal2018}).

To present the limitations of the two traditional types of orbital evolution we discuss the case of post-asymptotic giant branch (AGB) stars with intermediate orbital separations, such as the Red Rectangle (e.g., \citealt{VanWinckel2014}). The traditional evolutionary studies predict that the final orbital separation of the post-AGB star and the secondary star will either increase due to mass loss, or will substantially decrease if the giant and the secondary star enter a common envelope phase (e.g., \citealt{Izzardetal2010, Nieetal2012}). This expectation is in tension with some observed post-AGB intermediate binaries (post-AGBIBs) that have an intermediate orbital separation of $a \approx 1 \AU$, just inside the traditional gap (e.g., \citealt{Gorlovaetal2014, VanWinckeletal2014, Manicketal2017}). Observations (e.g., \citealt{Wittetal2009, Gorlovaetal2012, Gorlovaetal2015, Bollenetal2017, VanWinckel2017}) show that in probably most post-AGBIBs the more compact secondary star launches jets, likely wide jets \citep{Thomasetal2013, Bollenetal2017}, and that in most of them there is a circumbinary disc that testifies to a strong binary interaction. Similar evolution might take place for post-red giant branch (RGB) binaries (e.g., \citealt{Kamathetal2016}). 

Post-AGBIBs and other puzzles call for a third type of interaction. We adopt the view that the third type of interaction is the grazing envelope evolution (GEE). One of us \citep{Soker2017SNIIb} argued that the GEE can account for post-AGBIBs and similar objects, and for the progenitors of type IIb supernovae (those with very little hydrogen at explosion). 
  
The GEE is a newly proposed process \citep{Soker2015, Soker2016a} that is based on the removal of the outer giant envelope by jets that are launched by the secondary star. The secondary star grazes the envelope of the giant star while accreting mass from the envelope via an accretion disc. The accretion disc launches jets that efficiently remove the envelope in the vicinity of the orbit of the secondary star, and by that the jets prevent (or postpone) the system from entering the common envelope phase \citep{SabachSoker2015, Shiberetal2017, ShiberSoker2018}. During the GEE the orbital separation might decrease, it might not change much, or it might even increase. As well, during the long GEE the binary system can lose more mass from the second Lagrange point ($L_2$) beyond the compact secondary star. 
 
The mass transfer during the GEE is a combination of Roche-lobe overflow (RLOF) and accretion from an ambient gas (a Bondi-Hoyle-Lyttleton type of accretion). The jets operate in a negative feedback cycle \citep{Soker2016Rev}. If they remove too much gas, the accretion rate decreases and so does the jets' power (see, e.g., simulations by \citealt{MorenoMendezetal2017}). The removal of gas tends to increase the orbital separation, as well leading to a reduction in the accretion rate if tidal forces do not force the system to enter a common envelope. When the giant expands or tidal forces cause the secondary to approach the envelope again, the accretion rate and the power of the jets increase once again, and the jets remove more envelope mass. 
 
There is also a positive feedback in the launching of the jets. The jets remove gas with high energy and entropy from the accretion flow onto the secondary star. This reduces the pressure in the vicinity of the secondary star, hence enabling a high mass accretion rate \citep{Shiberetal2016, Staffetal2016MN}, as \cite{Chamandyetal2018} demonstrated in their recent numerical common envelope simulations. If jets do not remove high energy gas then the build-up of pressure in the vicinity of the secondary star that is embedded in a stellar envelope substantially reduces the accretion rate (e.g. \citealt{RickerTaam2012, MacLeodRamirezRuiz2015}). 

In the present study we incorporate the GEE into the \textsc{mesa binary} code \citep{Paxtonetal2015} in a simple phenomenological way, and follow the evolution of the orbital separation with and without the GEE. We describe the numerical setting in section \ref{sec:GEE}, we study the basic characteristics of the evolution in section \ref{sec:flow}, and we expand the parameter space in section \ref{sec:parameters}. Our summary is in section \ref{sec:summary}.

% ==========================================================
\section{MIMICKING THE GRAZING ENVELOPE EVOLUTION}
\label{sec:GEE}
% ==========================================================

In the present study we limit the parameter space to demonstrate the behavior of the GEE. We consider only one set of stellar masses and only circular orbits. We use the \textsc{binary} module of the \textsc{mesa} code (Modules for Experiments in Stellar Astrophysics, version 8845; \citealt{Paxtonetal2011,Paxtonetal2013,Paxtonetal2015}) to evolve a binary system composed of two stars with initial (on the main sequence) masses of $M_{1,0}=3.5 M_\odot$ (the mass donor) and $M_{2,0}=0.5 M_\odot$, and on a circular orbit. For the initial orbital separation we take several values in the range of $a=200$-$800 R_\odot$. In this first study of the GEE with \textsc{mesa binary} we consider the secondary star (the mass gainer) to be a point mass. The initial metallicity of the primary star is $Z=0.02$. Its initial equatorial rotation velocity is zero, though it can spin up due to tidal forces and then rotation is treated according to the `shellular approximation', where the angular velocity $\omega$ is assumed to be constant for isobars (e.g., \citealt{Meynet1997}). Orbital evolution due to tidal friction follows \cite{Hut1981}, with the timescales of \cite{Hurley2002} for convective envelopes. If the stars achieve contact, i.e., the separation equals the sum of their radii,
\begin{equation}
a=R_1+R_2, 
\label{eq:aR1R2}
\end{equation}
where in our simulations here $R_2=0$, the evolution is terminated. In some runs this condition is never met, and the evolution is terminated when $a$ and $R_1$ grow separate.
 
Mass transfer is treated in two different approaches, where in one we assume that jets launched by accretion onto the secondary star remove material from the envelope of the primary star, and in the second we assume that there are no jets. In the latter case, mass transfer due to RLOF is according to \cite{Kolb1990}, $\dot M_{\rm KR}$, multiplied by a factor of $\beta = 0.1$ in most runs, and $\beta=0.5$ in two runs. The explanation for this mass transfer factor of $\beta \simeq 0.1$ is as follows. \cite{Kolb1990} derive their mass transfer rate under the condition that the mass transfer proceeds on a thermal or nuclear timescale. This expression perhaps overestimates the mass transfer rate if the evolution is on a time shorter than the thermal timescale, as in the present study. When the mass transfer is on a dynamical time scale, the envelope does not have time to adjust itself. Its expansion speed is limited to the sound speed. On the surface of a cool star the sound speed is $\approx 10 \km \s^{-1}$, smaller than the Keplerian speed of the companion, $\approx 20$-$50 \km \s^{-1}$ in our cases. We run two cases with a moderate reduction by only a factor of 2, i.e., $\beta=0.5$, and as we discuss later, the GEE still works. So the value of $\beta$ is not a main issue. In a forthcoming study we will use $\beta=1$, but that will require a smoothing procedure to the turning on and off the jet activity, as in the present study it turns on to, and turn off from, maximum activity power abruptly.

We apply here our consideration that when mass transfer is rapid (on a shorter timescale than the thermal timescale), a different approach might be needed. Therefore, while in \textsc{mesa binary} the mass transfer rate in cases of RLOF is according to \cite{Kolb1990}, in the case of a red giant donor (RGB and AGB stars) where no jets are launched we take the mass accretion rate $M_{\rm acc,No-Jet}$ to be equal to the mass transfer rate $\dot M_{\rm t,No-Jet}$, with both being much smaller than the coded prescription
\begin{equation}
\dot M_{\rm acc,No-Jet} = \dot M_{\rm t,No-Jet} = \beta \dot M_{\rm KR}, 
\label{eq:KR01}
\end{equation}
with $\beta \ll 1$. The present value of $\beta=0.1$ (except for two runs for which $\beta=0.5$) is somewhat arbitrary, but it allows us to demonstrate the behavior of the GEE. 
    
Jets that are launched by the secondary star can remove mass not only from the envelope of the primary star, but also from the acceleration zone of the primary wind. For that and for numerical reasons, we assume that jets are launched and remove mass from the primary star when the separation is below a threshold that is somewhat larger than the radius of the primary photosphere. Our condition for the operation of jet-induced mass removal is 
\begin{equation}
a<f_\mathrm{GEE} \left(R_1+R_2\right),
\label{eq:fGEE}
\end{equation}
where here $R_2=0$. We use the values of $f_\mathrm{GEE}=1.03, 1.05, 1.1$ in the different runs. When the condition of equation (\ref{eq:fGEE}) is met, the mass transfer in \textsc{mesa binary} is enhanced by a factor we denote as $\eta$, with $90\%$ of the material being ejected and $10 \%$ being accreted. The expressions for mass accretion and mass ejection rates during the active phase of jets read 
\begin{equation}
\dot M_{\rm acc,Jet} = 0.1 \eta \dot M_{\rm t,No-Jet}, 
\label{eq:acc}
\end{equation}
and
\begin{equation}
\dot M_{\rm eject,Jet} = 0.9 \eta \dot M_{\rm t,No-Jet}, 
\label{eq:eject}
\end{equation}
respectively. We use values of $\eta=2, 3, 5$. This straightforward recipe is our way of mimicking the ejection of envelope material from the primary star by jets that are launched by the mass-accreting secondary star. Additionally, we reduce the time step when the jet activity begins, by setting the \textsc{mesa} variable \texttt{varcontrol\char`_target} to $10^{-5}$ (instead of the default value, $10^{-4}$). Whenever the condition of equation (\ref{eq:fGEE}) is not satisfied, mass transfer is according to equation (\ref{eq:KR01}).

We will show later that in many runs the jet activity turns on and off for tens of times. Therefore, during the several years of the intermittent jet activity, when the jets are off all the transferred mass is accreted, as given by equation (\ref{eq:KR01}). The result is that integrated over time, the secondary star accretes much more than $10\%$ of the mass lost by the primary star.

% ==========================================================
\section{EVOLUTION OF THE ORBITAL SEPARATION}
\label{sec:flow}
% ==========================================================

% =======================================
\subsection{Preventing the common envelope phase}
\label{subsec:prevent}
% =======================================

We start by presenting the properties and behavior of two runs, one with jets and one without. In sections \ref{subsec:postpone} and \ref{sec:parameters} we describe the results of our simulations with some different parameters. 
 
The cases we study here have the following properties. The initial masses of the two stars are $M_{1,0}=3.5 M_\odot$ and $M_{2,0}=0.5 M_\odot$, as in all simulations to follow. The initial orbital separation is $a_0=200 R_\odot$, and the reduced RLOF mass transfer rate parameter is $\beta=0.1$ (eq. \ref{eq:KR01}). In one simulation, termed NoJets200, we do not launch any jets. In the second run, termed Jet200-5-1.1, we assume that jets are launched by the secondary star when the condition of equation (\ref{eq:fGEE}) is met for $f_{\rm GEE}=1.1.$, namely, we take the mass accretion rate and mass ejection according to equations (\ref{eq:acc}) and (\ref{eq:eject}) and with $\eta=5$. 

\begin{figure*}
\centering
\includegraphics[trim=2.0cm 0.0cm 0.0cm 0.0cm,clip=true,width=1.2\textwidth]{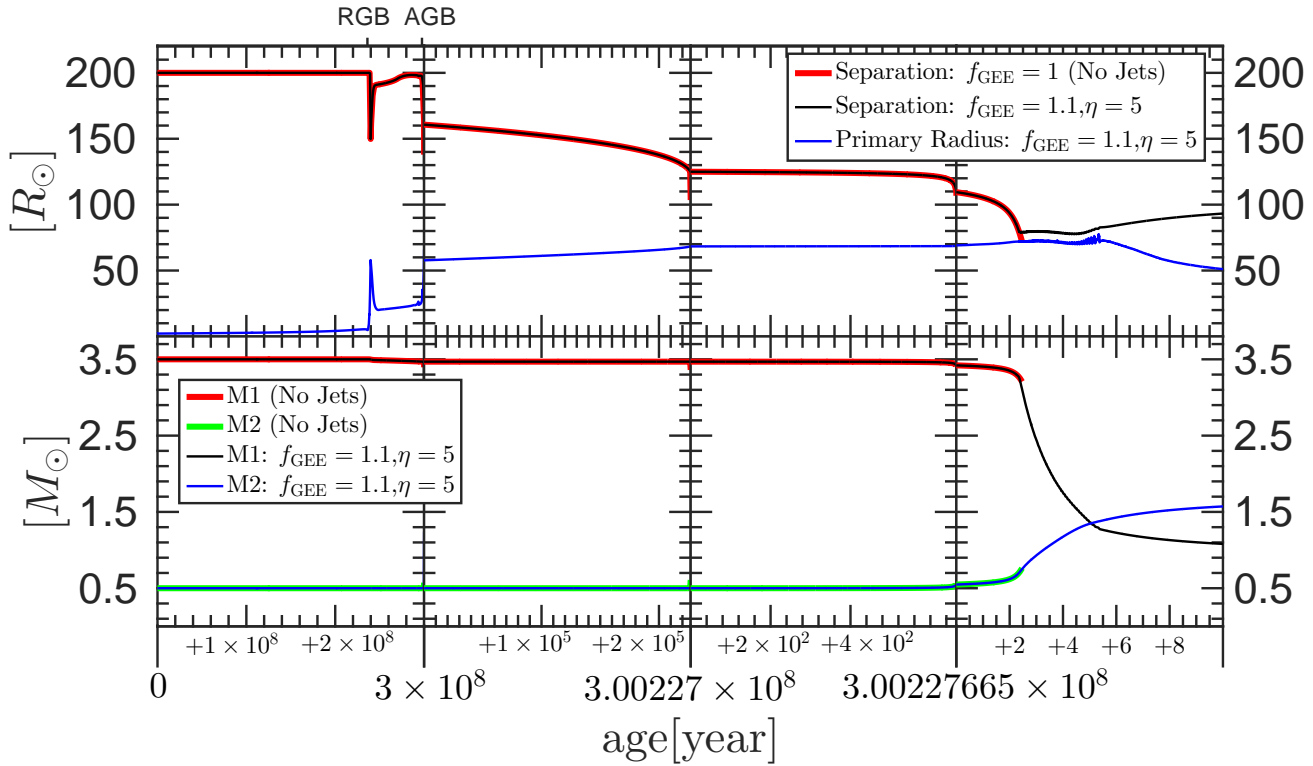}
\vskip -1.10 cm
\caption{Evolution from the zero age main sequence to the post-jet activity phase of a binary system with initial masses of $M_{1,0}=3.5 M_\odot$ and $M_{2,0}=0.5 M_\odot$ and an initial separation of $a_0=200 R_\odot$. Note that there are four time segments, each with a different scale. The lower larger numbers give the time in years from the zero age main sequence, while the smaller numbers closer to the axis give the extra time from the beginning of the time segment. All numbers are in years. The thick red line in the upper panel shows the orbital separation for run NoJets200 where jet activity does not take place. The system enters a common envelope phase and the calculation is terminated. The thin black line presents the orbital separation for run Jet200-5-1.1, where jet activity takes place when $a<f_\mathrm{GEE} R_1$, with $f_\mathrm{GEE}=1.1$. The thin blue line shows the radius of the primary star for the run with jets. The bottom panel shows the evolution of the stellar masses as indicated in the inset. We note that when the jets are active the secondary star accretes only $10\%$ of the transferred mass (eq. \ref{eq:acc}). However, during the many periods when there is no jet activity the secondary star accretes all the transferred mass. This is the reason that, overall, the secondary star accretes more than $10 \%$ of the transferred mass.}
\label{fig:run94}
\end{figure*}

We present the results of these two simulations in Fig. \ref{fig:run94}. Note that we split the horizontal axis to four different segments, each with a different timescale, becoming shorter from left to right. As long as the ratio of orbital separation to primary radius is large there is no strong interaction between the stars. When the primary radius expands along its RGB, as marked on Fig. \ref{fig:run94} at a time of about $2.4 \times 10^8 \yr$, tidal interaction takes place and leads to angular momentum exchange between the orbital motion and the rotation of the primary star. The reduced orbital angular momentum results in a decrease in the orbital separation and the primary star spins up. As the primary star shrinks during its post-RGB phase, it transfers angular momentum back to the orbit and the orbital separation increases. The orbital decrease repeats itself as the primary star climbs the AGB.
 % FFFFFFFFFFFFFFFFFFFFFFFFFFFFFFFFFFFFFFFFFFFFFFFF
% FFFFFFFFFFFFFFFFFFFFFFFFFFFFFFFFFFFFFFFFFFFFFFFF
 
Because the primary star both expands and loses angular momentum in the wind, angular momentum is transferred from the orbital motion to the primary star, to maintain synchronization of the orbital motion with the primary spin. As a result of that the orbit shrinks. If we assume that jets are not launched, hence do not remove mass from the primary envelope, the system enters a common envelope phase, as depicted by the thick red line (orbital separation) in the upper panel of Fig. \ref{fig:run94} that reaches the blue line (primary radius). 
 
If we assume, on the other hand, that when $a \le f_{\rm GEE} R_1$, where here $f_{\rm GEE}=1.1$, jets are launched by the secondary star and remove mass from the envelope of the primary star, then the system avoids the common envelope phase and enters the GEE. The orbital separation of the GEE is depicted by the thin black line in the upper panel of Fig. \ref{fig:run94}. Since we assume that at the same time as the primary star transfers mass to the secondary star jets remove more mass from the primary envelope, the  mass of the primary star decreases quite rapidly (thin black line in the lower panel of Fig. \ref{fig:run94}). At the same time the mass of the secondary star increases (thin blue line in the lower panel). The primary star shrinks and the orbital separation increases. The binary system avoids the common envelope phase and the final orbital separation is about $0.5 \AU$. The system will evolve to become a post-AGBIB system.
 
\begin{figure}
\centering
\includegraphics[trim=1.0cm 0.0cm 0.0cm 0.0cm,clip=true,width=0.48\textwidth]{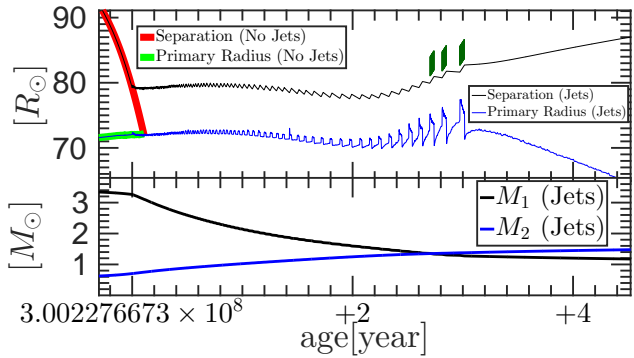}
\caption{Zooming in on the final years of our simulations that we present in Fig. \ref{fig:run94}, runs Jet200-5-1.1 and NoJets200, to reveal the differences between the runs with and without jet activity, respectively. Upper panel: The orbital separation $a$ in the case with (black line) and without (red line) jet-driven mass loss. The blue line and the green line are the primary radius $R_1$ in the cases with and without jets, respectively. The jets are turned on and off tens of times depending on the ratio $a/R_1$ according to equation (\ref{eq:fGEE}) with $f_{\rm GEE}=1.1$; the three green vertical lines mark the last three jet activity episodes. Lower panel: The masses of the two stars for the run with jets as indicated in the inset. % limits=$[3.002276671509725\times10^{8}\quad  3.0022767208\times10^8\quad66.88\quad 86.73]$ 
The jet-activity phase lasts for about 3 years. % $3.014year$ 
The time of the first large tick ($3.002276673 \times 10^8 \yr$) is the time when the jet activity starts, and each large tick marks two years. The tens of steps are due to numerical effects because the mass removal rate due to the jet activity is turned on instantaneously for its full value. This causes an increase in the orbital separation and a decrease in the primary radius, hence the jets are turned off by the condition of equation (\ref{eq:fGEE}). This repeats itself for close to a hundred times. In reality the line should be smooth.}
\label{fig:run94zoom}
\end{figure}

To better reveal the behavior of our numerical procedure that mimics the GEE, in Fig. \ref{fig:run94zoom} we zoom in on the jet-activity phase that lasts for about 3 years. After the jet activity starts, the orbital shrinkage slows down, and then increases somewhat, followed by further decrease and then a final increase. Although the companion is outside the envelope of the primary star (for numerical reasons), it is very close to the surface and the jets remove mass from the acceleration zone of the wind, which is also compatible with the GEE. In some of the other cases that we study in section \ref{sec:parameters} the companion is  closer to the primary surface when jets are launched. The tens of steps that are seen in the primary radius and orbital separation are numerical effects due to the jets turning on instantaneously and the limitation of the stellar model. So instead of a continuous operation of the jets, our modified version of the numerical code \textsc{mesa binary} turns the jets on and off for tens of times.   
% FFFFFFFFFFFFFFFFFFFFFFFFFFFFFFFFFFFFFFFFFFFFFFFF
% FFFFFFFFFFFFFFFFFFFFFFFFFFFFFFFFFFFFFFFFFFFFFFFF

The secondary star accretes about a solar mass in several years. To prevent the secondary star from expanding to large dimensions during the accretion process the jets must remove high entropy gas and a large amount of energy from the vicinity of the star \citep{Shiberetal2016}, in addition to removing angular momentum. At the very inner part of the thin accretion disk the magnitude of the gravitational energy of the gas is twice the  kinetic energy, and the internal energy is very small. Namely, the gas already obeys the virial relation on the surface of the star. By removing more energy from that region and ejecting the high entropy gas, the gas that eventually ends on the star has an average entropy lower than that in the outer layer of the secondary star. Some of the energy of the accreted gas that has high specific angular momentum is in rotation (e.g., \citealt{Kunitomoetal2017}), in our case rapid rotation, and this must be taken into account as well when calculating the response of the secondary star.
 
Even if the secondary star starts to inflate a light and large envelope, the jets might remove also part of this envelope (which is the gas with high entropy). The accretion disk geometry prevents the envelope to inflate, at least at early times, in the equatorial plane, and the envelope starts to inflate along the polar directions. The jets that are launched from the very inner parts of the accretion disk might remove envelope material along the poles further away. This suggested process must be checked with two-dimensional hydrodynamical simulation. We do expect the star to expand, but at early times more along the polar directions, even when stellar rotation is considered. For example, as a consequence of tripling its mass the secondary star might increase its radius by a factor of a few relative to a main sequence, namely by an order of magnitude or so. This estimate is based on the assumption that the jets carry a large fraction of the energy of the accreted mass, and hence the energy of the accreted mass is not high enough to allow expansion. In other words, the high accretion rate implies a dense disk that the jets `cool', and hence we expect the entropy of the accretion disk to be relatively low. This inflation will reduce the gravitational potential of the secondary star by a factor of about 3, still allowing for the launching of energetic jets. For the evolution described above to hold it is mandatory to eventually show that a main sequence star can accrete such a mass in several years and still launch energetic jets.

% =======================================
\subsection{Altering the evolution toward a common envelope}
\label{subsec:postpone}
% =======================================
 
\begin{figure*}
\centering
\includegraphics[trim=1.5cm 0.0cm 0.0cm 0.0cm,clip=true,width=1.2\textwidth]{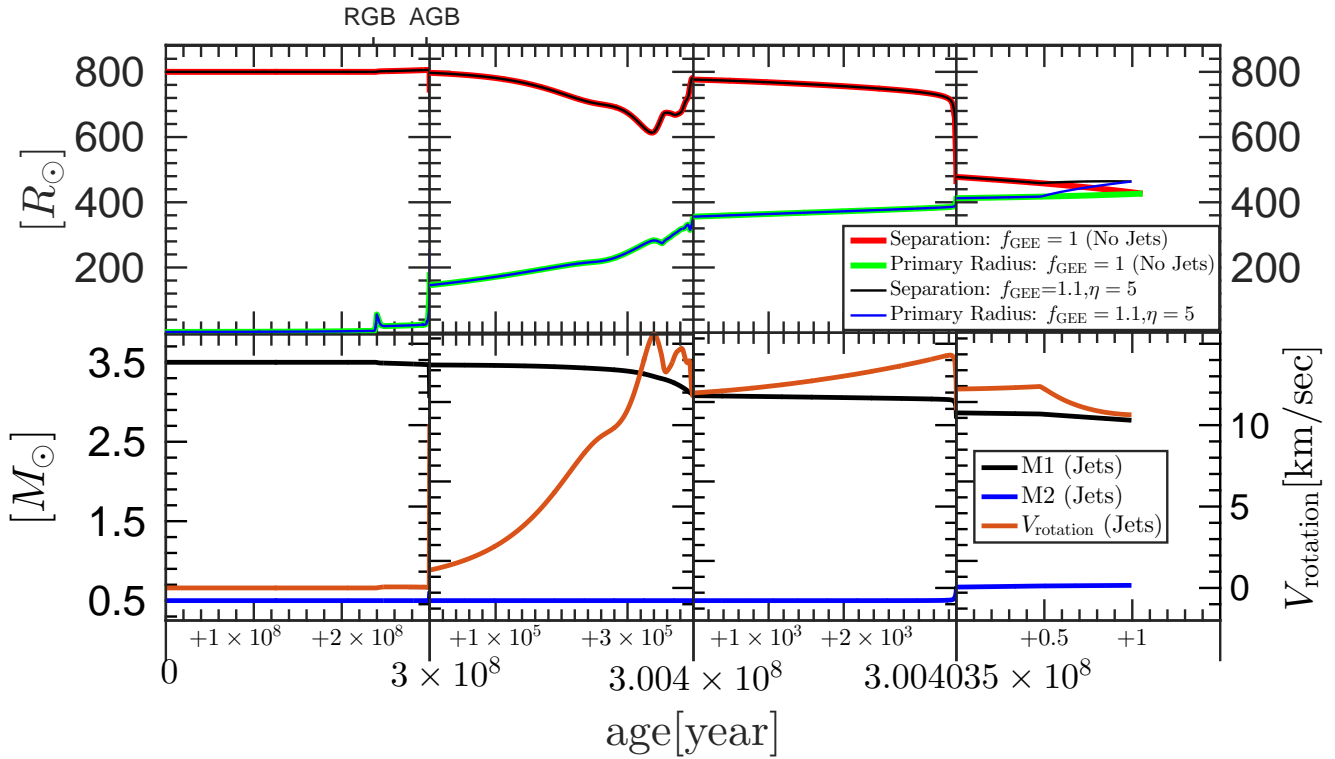}
\vskip -0.9 cm
\caption{Like Fig. \ref{fig:run94}, but with $a_0=800$, i.e., runs Jet800-5-1.1 and NoJets800, with the coding as given in the insets. In addition, the thick orange line in the lower panel depicts the rotational velocity on the primary equator for the case with jets, and the scale is on the right axis in the lower panel. In this case jet activity alters the entrance of the system to a common envelope phase (black line in upper panel) with respect to evolution without jets (thick red line), but does not prevent it. During the last phase of the evolution before the common envelope the jet activity is continuous, and hence the secondary star accretes $10 \%$ ($0.0079 M_\odot$) of the mass lost by the primary star ($0.079 M_\odot$).}
\label{fig:run107}
\end{figure*}

In some cases jet activity only alters or postpones the common envelope phase. We present one such case for the same initial stellar models and the same parameters of $f_{\rm GEE}=1.1$ in equation (\ref{eq:fGEE}) and $\eta=5$ in equation (\ref{eq:acc}) as before, but for an initial separation of $a_0=800 R_\odot$. We present the results in Fig. \ref{fig:run107}, that now includes also the rotational velocity on the equator of the primary star. There is no tidal interaction on the RGB phase because the orbital separation is too large for that. As the primary ascends the AGB, as seen in the blue/green line in the upper panel, tidal interaction takes place and the primary spins up (thick orange line in the lower panel) and the orbital separation decreases somewhat (thick red line in the upper panel). Note that because the first time segment spans $3 \times 10^8 \yr$ and the AGB evolution is rapid, there is a short vertical jump in the radius of the primary star and its rotation velocity at the very end of that part of the graph. The final spiraling-in phase toward the formation of a common envelope is seen in the last time segment of Fig. \ref{fig:run107}. 
% FFFFFFFFFFFFFFFFFFFFFFFFFFFFFFFFFFFFFFFFFFFFFFFF
% FFFFFFFFFFFFFFFFFFFFFFFFFFFFFFFFFFFFFFFFFFFFFFFF

\begin{figure}
\centering
\includegraphics[trim=0.5cm 0.0cm 0.0cm 0.0cm,clip=true,width=0.48\textwidth]{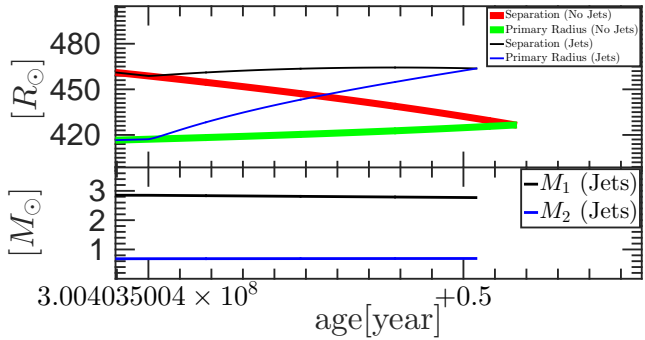}
\caption{Like Fig. \ref{fig:run94zoom} but zooming in on runs Jet800-5-1.1 and NoJets800 that we present in Fig. \ref{fig:run107}. Coding according to insets. }
\label{fig:run107zoom}
\end{figure}

We zoom in on the final year of our calculation in Fig. \ref{fig:run107zoom}. In the evolution without jets the radius of the primary star does not change much in the last year (thick green line) and the orbital separation decreases (thick red line). On the other hand, the removal of mass by jets causes the orbital separation to increase (black line), but as well to the expansion of the primary star (blue line) until the system enters a common envelope phase. Without jets the system enters the common envelope phase when the red line touches the green line, and with jets when the black line touches the blue line.   
% FFFFFFFFFFFFFFFFFFFFFFFFFFFFFFFFFFFFFFFFFFFFFFFF
% FFFFFFFFFFFFFFFFFFFFFFFFFFFFFFFFFFFFFFFFFFFFFFF

Although when jet activity takes place the system enters a common envelope phase, even faster than without jets, this is not the end of the story. As seen in Fig. \ref{fig:run107zoom}, the primary star expands by $\approx 10 \%$ of its radius in approximately its dynamical timescale of about several months. The envelope density in the outer regions drops in such a rapid expansion \citep{McleySoker2014}. Rather than entering a full common envelope phase, the jets launched by the secondary star can more efficiently now remove the primary envelope in the surroundings of the secondary star, as in the hydrodynamical simulations of \cite{Shiberetal2017}. So even in this case the jets might delay the entrance of the system to a full common envelope phase. In section \ref{sec:parameters} we will see cases where the jet activity delays the formation of the common envelope phase although it does not prevent it. 

% ==========================================================
\section{VARIATION OF PARAMETERS}
\label{sec:parameters}
% ==========================================================

To further reveal some of the possible outcomes of jet-induced mass removal we simulated several other cases. In Table \ref{Table:Runs} we list the different cases. In section \ref{sec:flow} we described the basic characteristics of the interaction by discussing two runs with and two runs without jet activity (figs. \ref{fig:run94}-\ref{fig:run107zoom}). We here present more cases, but we present only the relevant part where the evolution with jets departs from that without jets. 
%TTTTTTTTTTTTTTTTTTTTTTTTTTTTTTTTTTTTTTTTTTTTTTTTTTTTTTTTTTTTTT
\begin{table*}
\caption{Parameters and outcomes of numerical simulations}
\centering
\begin{threeparttable}
\begin{tabular}{cccccccccc}
\hline
Run        & $a_0$ ($R_\odot$) & $\eta$ & $f_\mathrm{GEE}$ & CEE & $M_{\rm 1f} (M_\odot)$ & $M_{\rm 2f} (M_\odot)$ & $a_f (R_\odot)$ & $P_f$(day) & Figs. \\
\hline
NoJets200  & $200$ & --    & $1.0$ & Yes& $3.20$ & $0.76 $ & $72$ & $35.5$& \ref{fig:run94}, \ref{fig:run94zoom} \\% 91 
NoJets300  & $300$ & --    & $1.0$ & Yes& $3.23$ & $0.73$ & $113$ & $69.9$& \ref{fig:run110zoom} \\% 106        
NoJets400  & $400$ & --    & $1.0$ & Yes& $3.24$ & $0.72$ & $151$ &  $108.0$&   \ref{fig:run109zoom}  \\% 105 
NoJets600  & $600$ & --    & $1.0$ & Yes& $3.22$ & $0.71$ & $231$ & $205.1$ &  \ref{fig:run108zoom} \\% 104 
NoJets800  & $800$ & --    & $1.0$ & Yes& $2.82$ & $0.70$ & $426$ & $542.9$ & \ref{fig:run107}, \ref{fig:run107zoom} \\% 103
Jet200-5-1.1 & $200$ & $5$ & $1.1$ & No & $0.87$ & $1.77$ & $113.7$ & $85$ & \ref{fig:run94}, \ref{fig:run94zoom} \\% 94
Jet200-5-1.05 & $200$& $5$& $1.05$ & No & $1.31$ & $1.37$ & $79$ & $50$ & \ref{fig:run98zoom}  \\% 98
Jet200-5-1.03 & $200$& $5$& $1.03$ & Yes& $1.43$ & $1.33$ & $75$ & $45.2$ & \ref{fig:run99zoom}  \\% 99
Jet200-3-1.1 & $200$ & $3$ & $1.1$ & No & $0.73$ & $1.89$ & $141$ & $120$ & \ref{fig:run100zoom} \\% 100 
Jet200-2-1.1 & $200$ & $2$ & $1.1$ & No & $0.73$ & $1.87$ & $143$ &$123$  & \ref{fig:run101zoom} \\% 101
Jet300-5-1.1 & $300$ & $5$ & $1.1$ & Yes& $1.97$ & $0.96$ & $127$ & $96.8$ & \ref{fig:run110zoom} \\% 110
Jet400-5-1.1 & $400$ & $5$ & $1.1$ & No & $0.89$ & $1.58$ & $260$ & $309$ & \ref{fig:run109zoom} \\% 109 
Jet600-5-1.1 & $600$ & $5$ & $1.1$ & Yes& $2.55$ & $0.75$ & $271$ & $284.5$ & \ref{fig:run108zoom} \\% 108 
Jet800-5-1.1 & $800$ & $5$ & $1.1$ & Yes& $2.77$ & $0.69$ & $463$ & $620.5$ &  \ref{fig:run107}, \ref{fig:run107zoom} \\% 107 
NoJets200($\beta$)   & $200$ & $-$& $1$ &Yes& $3.20$ & $0.75$ & $72$ & $35.6$ & \ref{fig:run112zoom} \\% 112 
Jet200-5-1.1($\beta$)& $200$ & $5$& $1.1$ &No & $0.72$ & $2.19$ & $140$ & $112$ &\ref{fig:run112zoom} \\% 112 
\hline
\hline
\end{tabular}
\footnotesize
\label{Table:Runs}
\begin{tablenotes}
The list of parameters that we vary between the binary evolution models we run, and some final values. $a_0$ is the initial binary separation, $\eta$ is the effective mass transfer enhancement when jet activity takes place (eq. \ref{eq:acc} and eq. \ref{eq:eject}), and $f_\mathrm{GEE}$ determines the orbital separation for jet activity to start (eq. \ref{eq:fGEE}). In the last two runs marked with $(\beta$) we take $\beta=0.5$ in equation \ref{eq:KR01}; in all other runs $\beta=0.1$. The fifth column indicates whether the system enters or not a common envelope phase. We then list the masses of the primary and secondary stars, and the orbital separation and period, all at the termination point of the evolution. In all runs the initial masses of the two stars are $M_{1,0}=3.5 M_\odot$ and $M_{2,0}=0.5 M_\odot$, respectively.
\end{tablenotes}
\end{threeparttable}
\label{tab:M30}
\end{table*}
% TTTTTTTTTTTTTTTTTTTTTTTTTTTTTTTTTTTTTTTTTTTTTTTT
 
\begin{figure}
\centering
\includegraphics[trim=1.0cm 0.0cm 0.0cm 0.0cm,clip=true,width=0.49\textwidth]{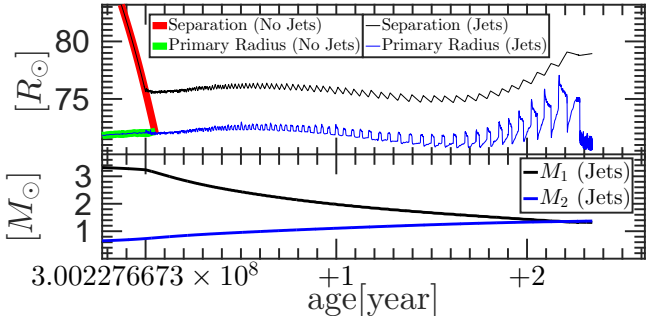}
\caption{Like Fig. \ref{fig:run94zoom}, but for run Jet200-5-1.05 with $f_{\rm GEE}=1.05$. The system avoids the common envelope phase. }
\label{fig:run98zoom}
\end{figure}

\begin{figure}
\centering
\includegraphics[trim=1.0cm 0.0cm 0.0cm 0.0cm,clip=true,width=0.49\textwidth]{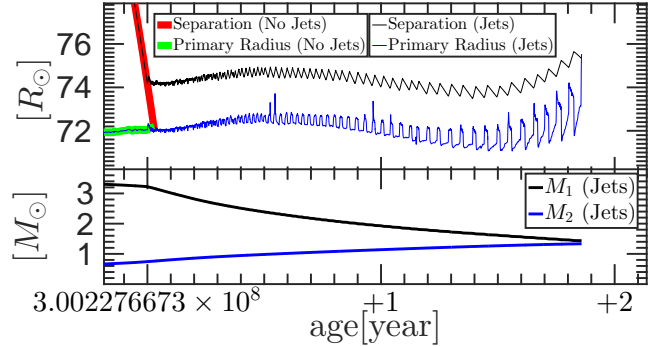}
\caption{Like Fig. \ref{fig:run94zoom}, but for run Jet200-5-1.03 with $f_{\rm GEE}=1.03$. The jet-driven mass removal postpones the common envelope phase, but does not prevent it. }
\label{fig:run99zoom}
\end{figure}

We first examine the role of the condition for the onset of jet activity. In run Jet200-5-1.1 that we presented in Fig.  \ref{fig:run94zoom} we took $f_{\rm GEE}=1.1$ in equation (\ref{eq:fGEE}). In Figs.  \ref{fig:run98zoom} and \ref{fig:run99zoom} we present the results for the cases with $f_{\rm GEE}=1.05$  and $f_{\rm GEE}=1.03$ , respectively. Namely, the removal of mass by jets starts only when the secondary star comes closer to the primary envelope. The other parameters are the same as in run Jet200-5-1.1. For $f_{\rm GEE}=1.05$ the system still avoids the common envelope phase. For the case when the activity starts only when the secondary is very close to the surface (run Jet200-5-1.03 $f_{\rm GEE}=1.03$; Fig. \ref{fig:run99zoom}), the jet activity postpones the onset of the common envelope phase by about two years. After that time the envelope rapidly expands and engulfs the secondary star. The numerical code stops at that stage. However, as we discussed in section \ref{subsec:postpone}, even when the secondary star orbits inside the outskirts of the giant envelope it can launch jets and remove more envelope mass. In some cases it is expected that the jets remove enough envelope mass to cause the system to exit the common envelope. We cannot treat this process here.
% FFFFFFFFFFFFFFFFFFFFFFFFFFFFFFFFFFFFFFFFFFFFFFFF
% FFFFFFFFFFFFFFFFFFFFFFFFFFFFFFFFFFFFFFFFFFFFFFFF
% FFFFFFFFFFFFFFFFFFFFFFFFFFFFFFFFFFFFFFFFFFFFFFFF
% FFFFFFFFFFFFFFFFFFFFFFFFFFFFFFFFFFFFFFFFFFFFFFFF

\begin{figure}
\centering
\includegraphics[trim=1.0cm 0.0cm 0.0cm 0.0cm,clip=true,width=0.49\textwidth]{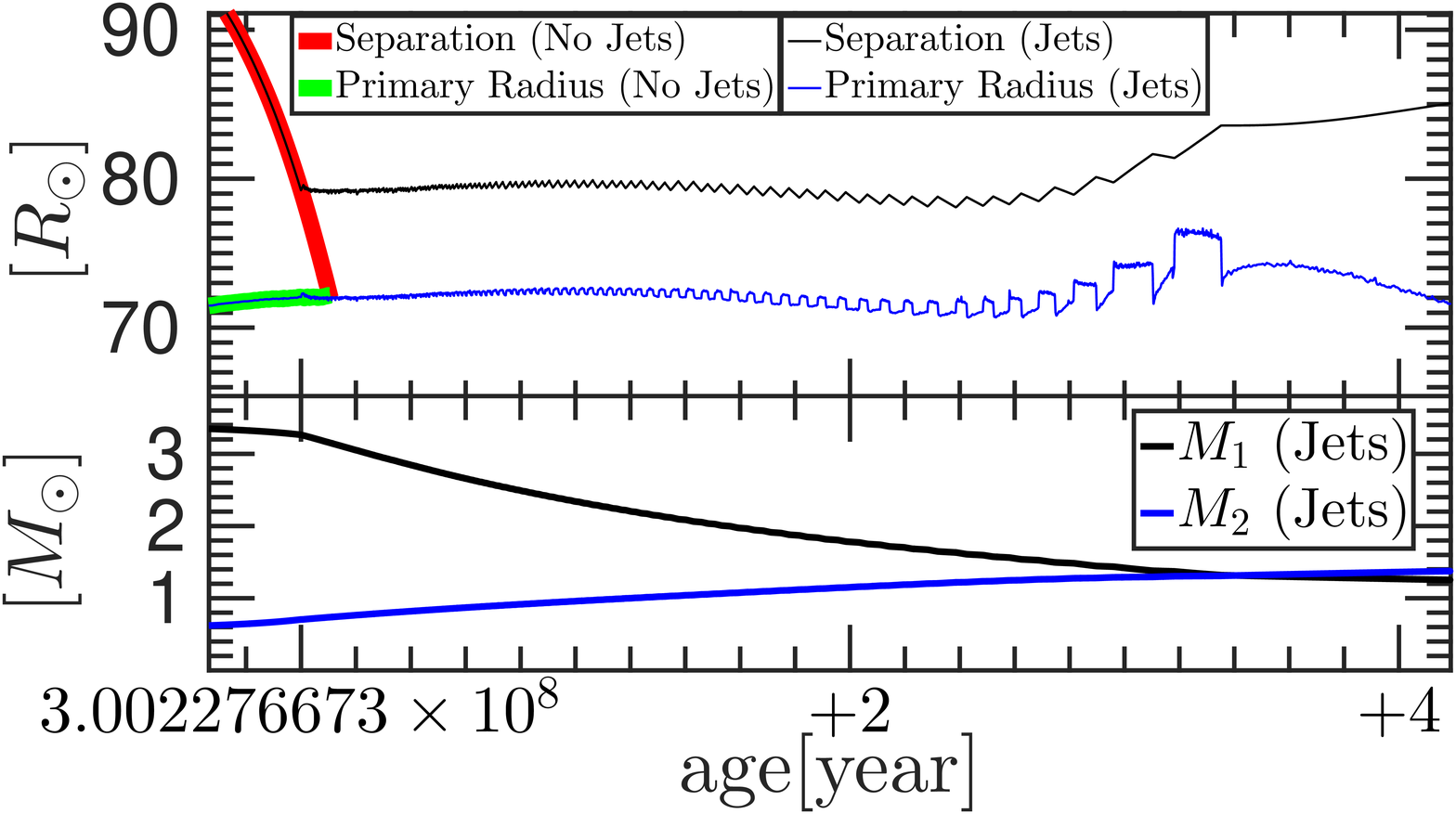}
\caption{Like Fig. \ref{fig:run94zoom}, but for run Jet200-3-1.1 with $\eta=3$.  }
\label{fig:run100zoom}
\end{figure}

\begin{figure}
\centering
\includegraphics[trim=1.0cm 0.0cm 0.0cm 0.0cm,clip=true,width=0.49\textwidth]{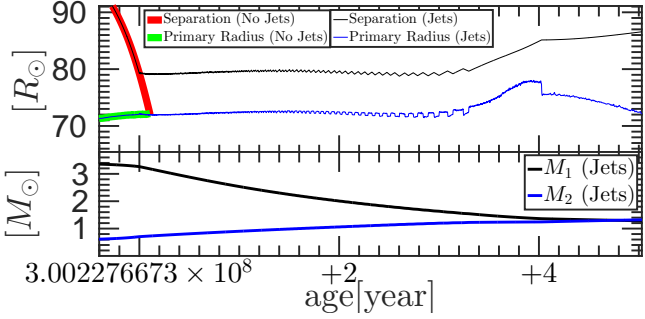}
\caption{Like Fig. \ref{fig:run94zoom}, but for run Jet200-2-1.1 with $\eta=2$. }
\label{fig:run101zoom}
\end{figure}

We next examine the case with a moderate enhancement of the mass transfer rate and the mass loss rate, namely, $\eta=3$ and $\eta=2$ instead of $\eta=5$ in equations (\ref{eq:acc}) and (\ref{eq:eject}). All the other parameters are as in run Jet200-5-1.1. We present the final evolution of these runs in Fig. \ref{fig:run100zoom} and Fig. \ref{fig:run101zoom}. We see that even with more moderate values of $\eta$ the binary system avoids the common envelope phase, and will enter the post-AGB phase at an intermediate orbital separation (i.e., a post-AGBIB). As expected, the evolution becomes longer as the value of $\eta$ decreases. 
% FFFFFFFFFFFFFFFFFFFFFFFFFFFFFFFFFFFFFFFFFFFFFFFF
% FFFFFFFFFFFFFFFFFFFFFFFFFFFFFFFFFFFFFFFFFFFFFFFF
% FFFFFFFFFFFFFFFFFFFFFFFFFFFFFFFFFFFFFFFFFFFFFFFF
% FFFFFFFFFFFFFFFFFFFFFFFFFFFFFFFFFFFFFFFFFFFFFFFF

\begin{figure}
\centering
\includegraphics[trim=0.5cm 0.0cm 0.0cm 0.0cm,clip=true,width=0.49\textwidth]{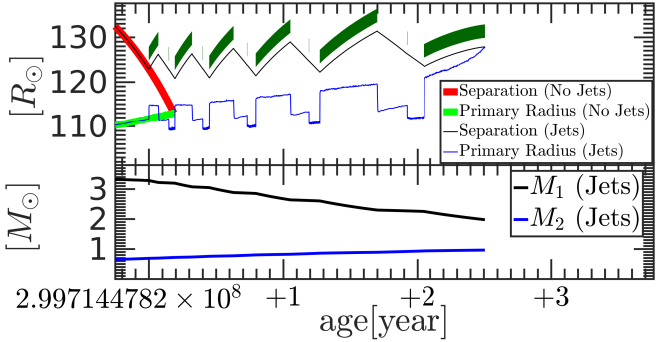}
\caption{Like Fig. \ref{fig:run94zoom}, but for runs NoJets300 and Jet300-5-1.1 with $a_0=300$. The green areas represent the time when the jet-induced mass removal is on (as before, according to the condition of eq. \ref{eq:fGEE}). }
\label{fig:run110zoom}
\end{figure}

\begin{figure}
\centering
\includegraphics[trim=0.5cm 0.0cm 0.0cm 0.0cm,clip=true,width=0.49\textwidth]{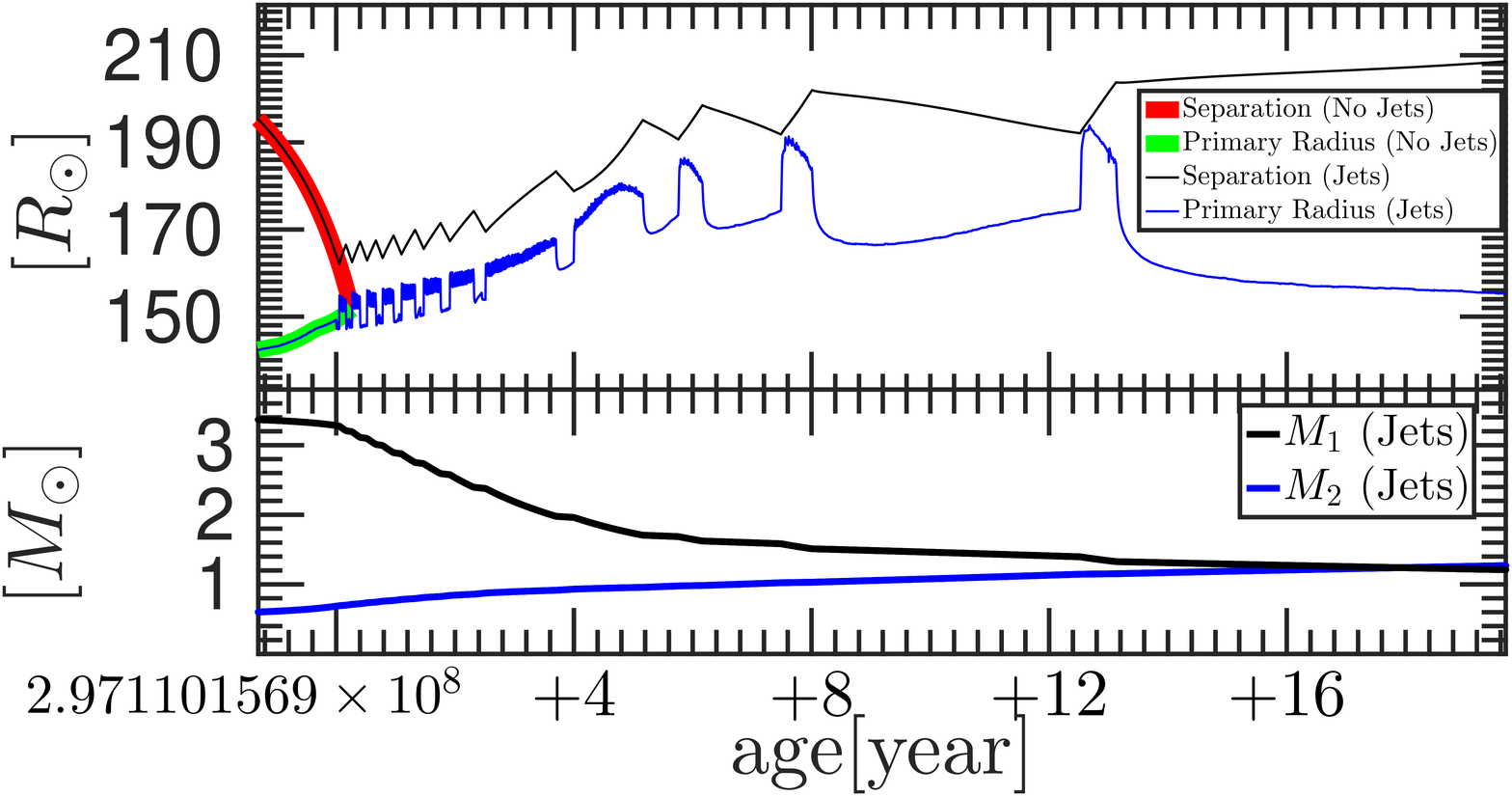}
\caption{Like Fig. \ref{fig:run94zoom}, but for runs NoJets400 and Jet400-5-1.1  with $a_0=400$.  }
\label{fig:run109zoom}
\end{figure}

\begin{figure}
\centering
\includegraphics[trim=0.5cm 0.0cm 0.0cm 0.0cm,clip=true,width=0.49\textwidth]{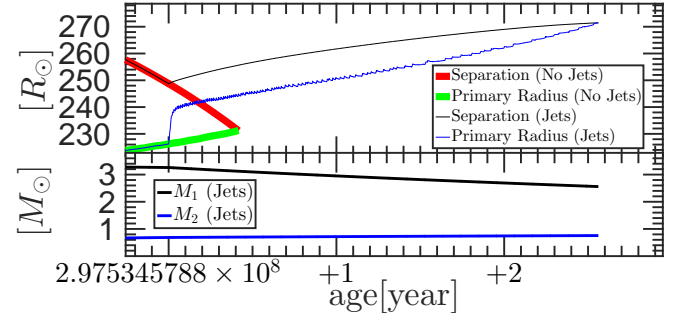}
\caption{Like Fig. \ref{fig:run94zoom}, but for runs NoJets600 and Jet600-5-1.1 with $a_0=600$.  }
\label{fig:run108zoom}
\end{figure}

We also examine the role of the initial orbital separation, holding the other parameters as in run Jet200-5-1.1 that we studied in section \ref{subsec:prevent}. In section \ref{subsec:postpone} we already presented the case with $a_0=800 R_\odot$. In Figs. \ref{fig:run110zoom}, \ref{fig:run109zoom} and \ref{fig:run108zoom} we present the cases for $a_0=300 R_\odot$, $a_0=400 R_\odot$ and $a_0=600 R_\odot$, respectively. The results are quite interesting when the initial orbital separation increases. For $a_0=300 R_\odot$ (fig. \ref{fig:run110zoom}) and $a_0=600 R_\odot$ (fig. \ref{fig:run108zoom}) a final expansion of the giant brings the system to enter a common envelope phase. This is similar to the final expansion with $a_0=800 R_\odot$ that we show in Fig. \ref{fig:run107zoom}. For $a_0=400 R_\odot$ (fig. \ref{fig:run109zoom}) the system almost enters a common envelope phase, but eventually avoids it. This suggests that in some cases even systems that do enter a common envelope phase might exit from it and resume the GEE, if the jets efficiently remove envelope mass when the secondary star orbits inside the outskirts of the giant envelope.
% FFFFFFFFFFFFFFFFFFFFFFFFFFFFFFFFFFFFFFFFFFFFFFFF
% FFFFFFFFFFFFFFFFFFFFFFFFFFFFFFFFFFFFFFFFFFFFFFFF
% FFFFFFFFFFFFFFFFFFFFFFFFFFFFFFFFFFFFFFFFFFFFFFFF
% FFFFFFFFFFFFFFFFFFFFFFFFFFFFFFFFFFFFFFFFFFFFFFFF
% FFFFFFFFFFFFFFFFFFFFFFFFFFFFFFFFFFFFFFFFFFFFFFFF
% FFFFFFFFFFFFFFFFFFFFFFFFFFFFFFFFFFFFFFFFFFFFFFFF

\begin{figure}
\centering
\includegraphics[trim=0.5cm 0.0cm 0.0cm 0.0cm,clip=true,width=0.49\textwidth]{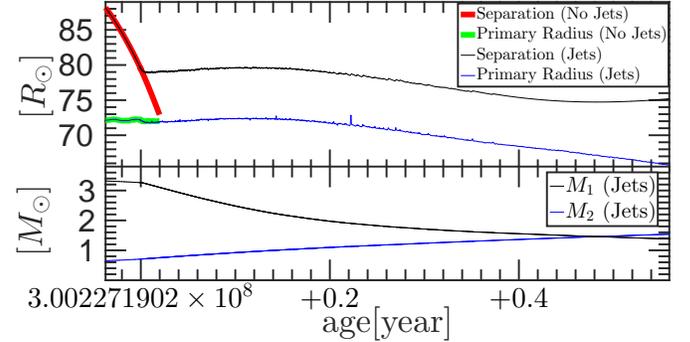}
\caption{Like Fig. \ref{fig:run94zoom}, but for runs NoJets200($\beta$) and Jet200-5-1.1($\beta$) with $\beta=0.5$ instead of $\beta=0.1$ as in all other runs, where $\beta$ is the reduction factor of the RLOF mass transfer rate.}
\label{fig:run112zoom}
\end{figure}

We end our sampling of the parameter space by taking $\beta=0.5$ in equation (\ref{eq:KR01}) rather than $\beta=0.1$ as in all the previous runs. Namely, we now have only a modest reduction of mass transfer rate in \textsc{mesa binary} which follows \cite{Kolb1990}. We present the results in Fig. \ref{fig:run112zoom}. We find that even for this moderate reduction in mass transfer rate, the mass removal by jets under our assumptions prevents the system from entering the common envelope phase. 
% FFFFFFFFFFFFFFFFFFFFFFFFFFFFFFFFFFFFFFFFFFFFFFFF
% FFFFFFFFFFFFFFFFFFFFFFFFFFFFFFFFFFFFFFFFFFFFFFFF
 
The final orbital separation $a_f$ and orbital period $P_f$ of our modeled systems that do not enter a common envelope phase are in the range of $\approx 80$-$260 R_\odot$ and $\approx 50$-$310 ~\mathrm{day}$, respectively (listed in Table \ref{tab:M30}). Post-AGBIBs have orbital periods in the general range of $\approx 100$-$3000 ~\mathrm{day}$ (e.g., \citealt{VanWinckel2007}). These systems are candidate post-GEE binaries (see section \ref{sec:intro}). The binary models we study here might explain some of the short orbital periods of these systems. Carbon-enhanced metal poor stars in binaries have an orbital period of $\approx 300$-$2000 ~\mathrm{day}$ (e.g., \citealt{Hansenetal2016}). These can also be systems that experienced a GEE. \cite{Polsetal2003} noted that traditional evolutionary models cannot account for the eccentricity and orbital periods of barium stars, for which the orbital periods are in the range of $\approx 80$-$10^4 ~\mathrm{day}$ (and for those with circular orbits $\approx 80$-$2000 ~\mathrm{day}$), and their masses are in the range $1$-$2.5 M_\odot$. Some of these systems could also have experienced a GEE.

Our systems overlap with the shorter periods of these systems. Systems that will end with larger final orbital separations might have a secondary star needing to accrete much less mass than in the present study. Nonetheless, the systems we study here might account for the general properties of  some of the short orbital period binary systems mentioned above.

We can summarize our main findings from this and the previous section as follows. Under our assumption (based on the numerical simulations of \citealt{Shiberetal2017} and \citealt{ShiberSoker2018}) that jets launched by the secondary star can remove substantial mass from the giant envelope, in many cases that we have simulated with \textsc{mesa binary}, but not all cases, the system can avoid the common envelope phase. To prevent the common envelope phase, the secondary star accretes a relatively large amount of mass, $\approx 1M_\odot$. We further discuss this in the next section. 
 
% ==========================================================
\section{SUMMARY}
\label{sec:summary}
% ==========================================================

The immediate goal of our paper is to present some characteristics of the GEE. We are not aiming in this first study of the GEE with \textsc{mesa binary} at specific binary systems. Our future goals are to study specific types of systems, like observed post-AGBIBs and the progenitors of Type IIb supernovae, and eventually to find a prescription to incorporate the GEE into population synthesis numerical codes.  

To reach our goal we have characterized the binary interaction with several parameters. These are the reduction from the mass transfer rate (which follows \citealt{Kolb1990}) of \textsc{mesa binary}, $\beta$ (eq. \ref{eq:KR01}), the orbital separation to primary radius ratio below which jet activity takes place, $f_{\rm GEE}$ (eq. \ref{eq:fGEE}), and the enhanced mass transfer rate during the jet activity phase, $\eta$ (eq. \ref{eq:acc} and eq. \ref{eq:eject}). We mimic jet-driven mass removal by taking the enhanced mass transfer (through $\eta$) and taking a fraction of 0.9 of the transferred mass to be lost from the system. 

In Table \ref{Table:Runs} we summarize the different cases we have simulated and list the figures where we present the evolution of each case. In all cases the initial masses of the primary and secondary stars were $M_{1,0}=3.5 M_\odot$ and $M_{2,0}=0.5 M_\odot$, respectively. We have found that under our assumptions and the way we mimic the jet-driven mass loss, in many cases, but not all, the jet-driven mass loss process can prevent the common envelope phase altogether. The system experiences the GEE. In these cases the orbital separation does not change much from the time the jet activity starts. The system will turn into a post-AGBIB system. In other cases the system does enter a GEE for a short time, but then transitions into a common envelope phase. 

One of our assumptions that needs further study is that the secondary star can accrete $\approx 1 M_\odot$ and still launch jets. It might be that it expands and the system does enter a common envelope phase. On the other hand, we expect that dynamical effects, such as mass loss from the second Lagrange point beyond the secondary star, will remove more mass than the jets alone remove. As well, even when the secondary star does enter the envelope we expect that jets continue to remove envelope mass, and in some cases this will be efficient enough so that the secondary continues to graze the giant envelope and the system never enters a common envelope phase. Also, in this study we have started with a relatively large mass ratio, $M_{1,0}/M_{2,0}=7$. It might be that in most cases the systems that experience the GEE and avoid the common envelope phase have a somewhat lower ratio, e.g., $M_{1,0}/M_{2,0} \simeq 3$-$5$.  

Overall, due to our simplifying assumptions we cannot yet present the parameter space for the GEE. More studies are required to allow the GEE to be incorporated into population synthesis numerical codes. Nonetheless, our study does suggest that the GEE is important and should be already considered in analyzing results of population synthesis studies of binary systems and in proposing scenarios for some puzzling evolved binary systems (see section \ref{sec:intro}). 

\section*{Acknowledgments}

We thank an anonymous referee for helpful comments. We acknowledge support from the Israel Science Foundation and a grant from the Asher Space Research Institute at the Technion. AG thanks the support of the Blavatnik Family Foundation.

\ifmnras
\bibliographystyle{mnras}

\label{lastpage}
\end{document}